# Harnessing Quantum Computing for Energy Materials: Opportunities and Challenges

*Seongmin Kim[1*], In-Saeng Suh[1], Travis S. Humble[2], Thomas Beck[1], Eungkyu Lee[3*] and Tengfei Luo[4*]*

AUTHOR ADDRESS

[1]National Center for Computational Sciences, Oak Ridge National Laboratory; Oak Ridge, Tennessee, 37830, United States.

[2]Quantum Science Center, Oak Ridge National Laboratory; Oak Ridge, Tennessee, 37830, United States.

[3]Department of Electronic Engineering, Kyung Hee University; Yongin-Si, Gyonggi-do, 17104, Republic of Korea.

[4]Department of Aerospace and Mechanical Engineering, University of Notre Dame; Notre Dame, Indiana, 46556, United States.

AUTHOR INFORMATION

**Corresponding Author**

Seongmin Kim – National Center for Computational Sciences, Oak Ridge National Laboratory; Oak Ridge, Tennessee, 37830, United States. E-mail: kims@ornl.gov

Eungkyu Lee – Department of Electronic Engineering, Kyung Hee University; Yongin-Si, Gyonggi-do, 17104, Republic of Korea. E-mail: eleest@khu.ac.kr

Tengfei Luo – Department of Aerospace and Mechanical Engineering, University of Notre Dame; Notre Dame, Indiana, 46556, United States. E-mail: tluo@nd.edu

ABSTRACT

Developing high-performance materials is critical for diverse energy applications to increase efficiency, improve sustainability and reduce costs. Classical computational methods have enabled important breakthroughs in energy materials development, but they face scaling and time-complexity limitations, particularly for high-dimensional or strongly correlated material systems. Quantum computing (QC) promises to offer a paradigm shift by exploiting quantum bits with their superposition and entanglement to address challenging problems intractable for classical approaches. This perspective discusses the opportunities in leveraging QC to advance energy materials research and the challenges QC faces in solving complex and high-dimensional problems. We present cases on how QC, when combined with classical computing methods, can be used for the design and simulation of practical energy materials. We also outline the outlook for error-corrected, fault-tolerant QC capable of achieving predictive accuracy and quantum advantage for complex material systems.

TOC GRAPHICS

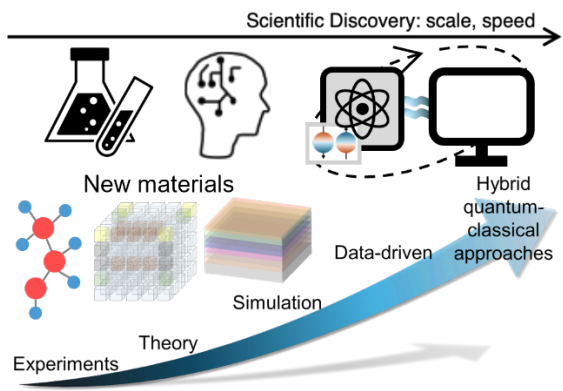

Energy consumption has reached unprecedented levels, driven by the global population growth, economic development, industry and technology expansion.[1] This trend necessitates more rapid development of energy materials that are efficient, durable, inexpensive and sustainable,[2] to enable key technologies such as passive cooling, lightweight transportation, advanced catalysts, high-performance batteries, and efficient photovoltaic and thermoelectric systems.[3-8] Developing new energy materials through empirical experimental screening is often time-consuming and resource-intensive.[9] Analytical models can reduce the trial-and-error efforts for some engineering objectives, but they typically require substantial time to develop and validate. With the advancement of computational power, numerical methods like density functional theory (DFT) and molecular dynamics (MD) become capable of predicting various material properties to supplement, and sometimes guide experiments.[10] However, simulating complex multiscale material systems can still be insurmountable for these methods. Recently, data-driven approaches, particularly machine learning (ML), have been extensively explored to accelerate high-throughput screening of candidate materials much more efficiently (Figure 1).[11]

Despite these advances, classical computational methods face fundamental limitations. Many problems in energy materials optimization involve vast design spaces that scale exponentially with the system size.[3, 12, 13] Such problems, including the configuration of photonic structures[3, 14, 15] or the design of high-entropy alloys,[16] are classified as non-deterministic polynomial (NP)-hard problems, and their solution becomes intractable for classical computers as problem complexity grows. Similarly, simulating quantum chemistry on classical computers can be challenging because the computational cost scales exponentially with the number of electrons and orbitals that need to be explicitly described in a material. Capturing electron correlation requires simplifications or partial representations, such as those used in coupled-cluster or configuration interaction methods, which quickly become infeasible for large, strongly correlated systems.[12, 13]

Quantum computing (QC) promises to enable a paradigm shift by exploiting quantum superposition and entanglement to represent and process information.[17, 18] In QC, the basic unit of information is quantum bit (qubit), which can exist in a superposition of states, unlike a classical bit that is either 0 or 1.[19] Multiple qubits can be entangled, and their quantum state spans a Hilbert space whose dimension grows exponentially with the number of qubits, allowing n qubits to encode $2^n$ complex numbers.[19-21] This exponential state space enables QC to represent and manipulate a large number of parameters efficiently, which is particularly advantageous for modeling complex, high-dimensional systems.[21] However, quantum states are fragile and susceptible to decoherence, the loss of quantum information over time due to interactions with the environment.[22] The duration over which a qubit maintains its quantum state, known as coherence time, fundamentally limits the depth and accuracy of quantum computations on current quantum hardware. In addition, the imperfect control of quantum gates (i.e., gate fidelity) introduces operational errors that accumulate as circuit depth increases (i.e., those requiring many quantum gate operations), further constraining the achievable accuracy of near-term QC.[23] Beyond gate-based QC architectures, adiabatic QC provides an alternative mechanism in which a problem is encoded as an Ising or QUBO Hamiltonian by expressing an objective function in terms of spin or binary variables. The quantum system then evolves slowly from an initial Hamiltonian to this problem Hamiltonian, such that its final ground state encodes the optimal solution, making adiabatic QC particularly well suited for combinatorial optimization.[24]

Although QC has the potential to explore combinatorial design landscapes and model strongly correlated systems more efficiently than classical methods,[25,26] current noisy intermediate-scale quantum (NISQ) devices are limited in both scale and accuracy due to the constraints (limited qubit counts, short coherence time, limited qubit connectivity and imperfect gate fidelity).[23,27] As a result, QC alone has not yet delivered significant advantages for problems at practical scales. Nonetheless, rapid progress in quantum hardware and algorithms suggests that QC may soon play a transformative role in practical problems, including energy materials design. Importantly, QC's impact has already emerged through hybrid quantum-classical methods, where quantum resources focus on bottleneck tasks such as combinatorial optimization or Hamiltonian simulation, while classical resources handle data preprocessing, feature extraction, and large-scale integration (Figure 1).[28,29] For instance, ML models can encode materials optimization problems into Hamiltonians that suits quantum algorithms, which can then be solved more efficiently with high-performance computing (HPC)-QC integrated systems.[29]

This perspective highlights the opportunities and challenges of harnessing QC for designing and simulating material systems, with particular emphasis on energy applications. We discuss recent algorithmic developments, explore the advantages of quantum approaches, and outline current challenges along with our perspectives. Finally, we provide a near- to long-term outlook for transitioning from current proof-of-concept demonstrations toward fault-tolerant QC capable of accelerating materials research at scale. By highlighting both limitations and potentials, we aim to provide a practical and forward-looking view of how QC may shape the future of energy materials development.

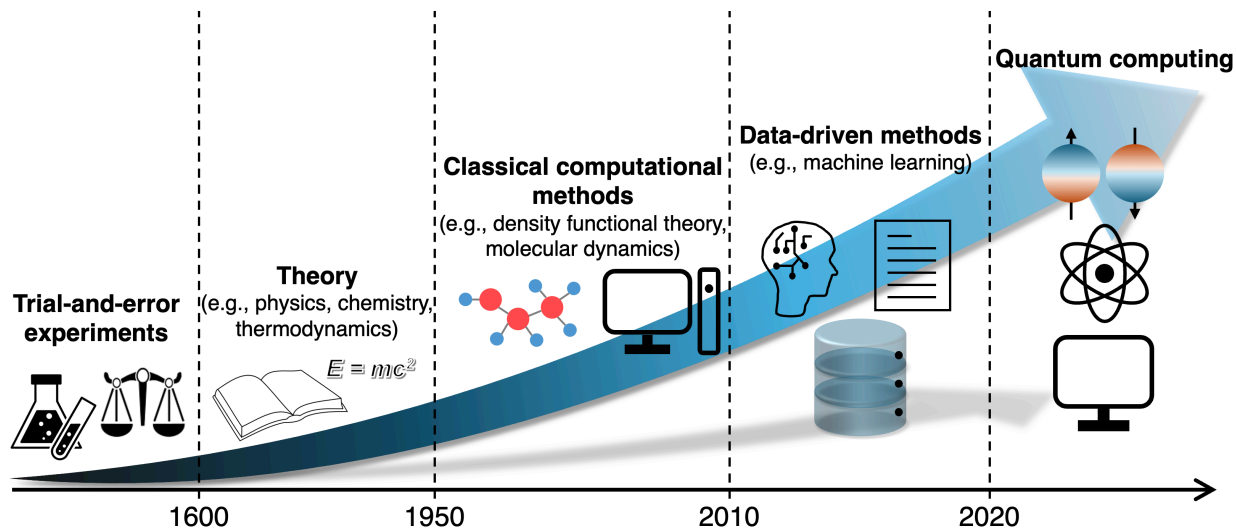

**Figure 1. Advancing scientific discovery through the evolution of methods.** Scientific discovery has advanced through successive paradigm shifts in methodology, moving from experiments to physics-based theory, to numerical simulation, to data-driven approaches, and now toward QC, each enabled by transformative advances in theories and technologies.

## Quantum Computing Methods

Unlike classical computing, which encodes information in deterministic binary bits (0 or 1), QC uses qubits that can exist in superpositions of states and become entangled, [27] and it processes quantum information with qubits through unitary quantum gates. By exploiting superposition and entanglement, QC can explore large Hilbert spaces by qubits, enabling more efficient exploration of many problems that may be intractable for classical algorithms, [30] such as large-scale combinatorial optimization tasks[3] or modeling strongly correlated quantum systems (Figure 2). [13]

Optimization problems encountered in energy materials discovery, such as the configuration of photonic structures[3, 14, 15] and high-entropy alloys[16], are NP-hard. QC can excel in solving these problems by leveraging quantum principles to efficiently explore the vast optimization spaces. In addition, QC enables new approaches to chemistry by encoding and manipulating quantum states in ways that mimic the natural evolution of quantum systems. [31] Electronic states in molecules are inherently quantum. Modeling these systems using classical methods relies on computationally expensive calculations or approximations that sacrifice either accuracy or scalability because the computational cost of treating electron grows exponentially with system size. [32] In contrast, QC promises to represent such systems faithfully through Hamiltonian simulation since n qubits naturally encodes the $2^n$-dimensional Hilbert space. This enables QC to capture many-electron entanglements without reducing or truncating the configuration space, potentially enabling accurate calculations of electronic properties with significantly reduced approximations. [26, 33-37] QC has been applied in early-stage demonstrations to study small molecules, such as LiH, $BeH_2$, and $H_6$.[38] Although these molecules are chemically simple, they represent fundamental building blocks of real materials as molecular units determine materials properties. Therefore, demonstrating QC on small molecules provides an important validation of the underlying quantum simulation principles, including Hamiltonian to qubit mapping, preparing correlated quantum states, and evaluating energy landscapes, which are expected to extend to complex materials.

However, extending these demonstrations to complex and large systems is still challenging due to current hardware limitations: (1) limited numbers of available qubits, and (2) restricted qubit connectivity, which constrain the size and complexity of molecular Hamiltonians that can be encoded. Dimensionality reduction strategies may be required for making early-stage QC useful in chemistry and materials science.[39] (3) Limited gate fidelity, which restricts circuit depth and accuracy, and (4) limited sampling rate, which imposes substantial measurement to obtain statistically reliable results. Continued advances in qubit coherence, connectivity, gate fidelity, quantum error mitigation/correction, and statistics of observed measurements will be essential to reaching chemical accuracy. Despite these barriers, rapid progress in quantum hardware and algorithms offers a promising pathway toward practical quantum advantage in materials.

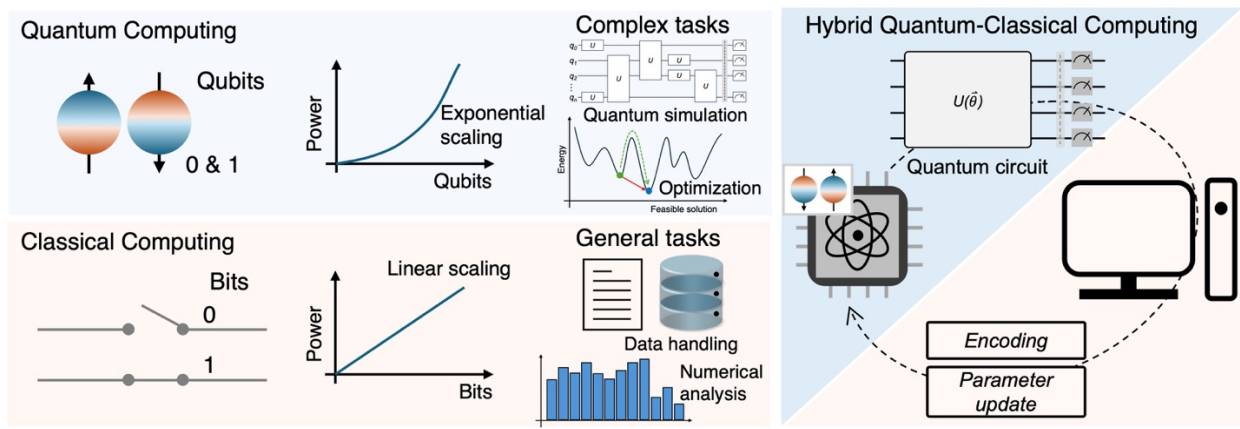

**Figure 2. Comparison between quantum and classical computing** in terms of fundamental units, computational scaling, and suitable use cases. Hybrid quantum-classical workflows leverage the complementary strengths of both paradigms.

For energy materials, two main QC technologies dominate, including adiabatic quantum computing and gate-based quantum computing.

*Adiabatic Quantum Computing*

Adiabatic quantum computing, often realized through quantum annealing, exploits the adiabatic theorem of quantum mechanics, which states that a quantum system will remain in its ground state if its Hamiltonian is varied sufficiently slowly. [24, 40] In quantum annealing, the problem of interest is encoded as a Hamiltonian, commonly expressed in a quadratic unconstrained binary optimization (QUBO) form, which is mathematically equivalent to an Ising model. The process begins with an initial Hamiltonian whose ground state is easy to prepare (e.g., uniform superposition of all possible states). The system then evolves gradually into the target problem Hamiltonian. If the evolution is sufficiently slow and the system avoids excitations due to noise, the final state is expected to correspond to the ground state of the Hamiltonian. [41]

Commercial quantum annealers, such as those developed by D-Wave Systems, can solve QUBO problems with thousands of qubits, despite limitations on connectivity and noise. Quantum annealers are well-suited for combinatorial optimization problems, which can be encoded or approximated as QUBOs. In material science, many optimization tasks can be approximated to quadratic functions, enabling their transformation into QUBO problems. Such problems are abundant in energy-related applications, such as configuration optimization, composition optimization, scheduling and logistics for energy systems. [24, 41] Although quantum annealing cannot address all scientific problems since only QUBO-type formulations can be directly solved, it represent a practical QC approach for large-scale and practical optimization tasks relevant to energy applications.

*Gate-Based Quantum Computing*

Gate-based QC operates on a more flexible model, using quantum circuits composed of unitary gates applied to qubits. [24] Hamiltonians representing quantum problems, such as those in quantum chemistry, can be mapped to quantum circuits, then QC can simulate this Hamiltonian to calculate its minimum eigenvalue and corresponding eigenstate. [42] In principle, gate-based QC is universal, meaning any quantum algorithms can be realized by a sequence of quantum gates, given sufficient qubits and error correction. [43] Quantum simulators implemented in classical computers can play an important role in testing quantum circuits under both noiseless and noise-included conditions, but their scalability is significantly limited by the resources required to simulate many qubit operations. Hence, simulating wide quantum circuits (i.e., circuits with a large number of qubits) is usually difficult on quantum simulators. In contrast, quantum hardware is not fundamentally limited by circuit width if sufficient qubits are available.

*Architecture*. There are several types of physical architectures for gate-based QC, each with distinct advantages and challenges. Across all platforms, quantum information is encoded in qubits, which can be implemented using different underlying technologies. Superconducting qubits offer fast gate speeds and good scalability potential, but they suffer from short coherence times. [44] Trapped-ion qubits feature exceptionally long coherence times and high-fidelity gates, though gate operations are relatively slow and scaling up large ion arrays is technically demanding. [45, 46] Photonic qubits enable room-temperature operation and efficient communication, but photon loss and universal two-qubit operations are key obstacles. [47, 48] Neutral atom qubits, manipulated by optical tweezers, allow for flexible connectivity and promising scalability, although gate precision still needs improvement. [49] Solid-state defect qubits, such as nitrogen-vacancy centers in diamond, offer long coherence and room-temperature operation, but scalable qubit coupling and control are not yet demonstrated. [48, 50] These QC architectures represent trade-offs between coherence, gate fidelity, speed, and scalability, and ongoing research is aiming to identify which platform, or hybrid combination, will realize fault-tolerant QC.

*Algorithms*. In the current NISQ era, the number of qubits is limited, and additional challenges such as short coherence times, gate errors, and restricted connectivity prevent the reliable execution of deep circuits. [23] To utilize the current gate-based QC for practical problems, a hybrid quantum-classical strategy is usually needed. Researchers have developed variational quantum algorithms (VQAs), which combine parameterized quantum circuits (ansatz) with classical optimizers. The quantum circuits prepare trial quantum states, while the classical optimizers adjust the variational parameters to minimize a cost function, such as the expectation value of a Hamiltonian. [51] With the current NISQ devices, VQAs have been successfully demonstrated for small systems, [38, 51, 52] but scaling to larger, more realistic materials is challenging due to hardware noise, limited qubit counts, restricted qubit connectivity, circuit depth constraints, and algorithmic constraints. Nevertheless, VQAs provide a promising foundation for tackling increasingly complex problems because qubits can natively represent and process information in exponentially large Hilbert spaces, enabling efficient exploration of the high-dimensional combinatorial optimization spaces and strongly correlated quantum systems. As quantum hardware and algorithms continue to advance, these capabilities are expected to extend toward more practical materials applications. Current key VQAs include:

(1) Quantum Approximate Optimization Algorithm (QAOA): QAOA has been widely used for combinatorial optimization problems such as materials structure design. [29] It alternates between applying problem Hamiltonian and mixing Hamiltonian, where each repetition of these operations is called a layer. The number of layers determines the circuit depth and controls the balance between accuracy and resource feasibility. Unlike quantum annealing, which uses an analog quantum approach based on continuous-time evolution, QAOA discretely approximates adiabatic evolution from Trotterization. [53] QAOA can be employed to optimize structural configurations of energy materials or scheduling and logistics for energy systems, which can be formulated as combinatorial optimization problems. However, the limited qubit counts and shallow circuit depths of current quantum hardware restrict the size of problems that standard QAOA can address. To overcome this limitation and enable larger-scale optimization, distributed QAOA has been proposed, leveraging HPC-QC integrated systems to distribute workloads across classical and quantum resources that operate in parallel. [29]

(2) Variational Quantum Eigensolver (VQE): VQE is widely used to estimate the ground state energies of molecular systems in material science. [18, 51] In this approach, a parameterized quantum circuit to prepare the quantum state, and its parameters are iteratively optimized using a classical optimizer to minimize the expectation value of the Hamiltonian. [34] Algorithmic developments for VQE continue to improve resource efficiency, and thus the practical calculations of complex molecules, such as SrVO$_3$, are within reach. [54]

(3) Adaptive Variants (ADAPT-QAOA, ADAPT-VQE): Adaptive methods dynamically build quantum circuits by iteratively adding gates or operators that most reduce the cost function. [38, 55] This results in more compact and problem-specific circuits compared to fixed-depth ansatzes. These adaptive methods can be utilized for combinatorial optimization[55] with potential relevance to material structure optimization and ground state energy estimation, [38] while reducing circuit depth and thereby enhancing scalability and robustness against noise on near-term quantum devices.

(4) QAOA-GPT: QAOA-GPT leverages generative pretrained models to automatically design quantum circuits tailored to specific problems. [56] It can generate problem-adapted ansatz structures, potentially reducing the effort required for ansatz engineering. QAOA-GPT can autonomously generate optimized quantum circuits for modeling material-specific Hamiltonians, greatly reducing circuit depth and accelerating materials discovery workflows.

ML methods can further enhance VQAs by guiding parameter initialization or enabling parameter transfer from related problem instances, providing better starting points for optimization. These strategies reduce the number of iterations needed, improve convergence to high-quality solutions, and increase the robustness and scalability of VQAs for larger or more complex problems. [57-59] In addition, ML models can be employed to mitigate or correct quantum errors, thereby improving the fidelity of quantum algorithms. [60, 61] Furthermore, recent advances in hybrid quantum-classical approaches, such as sample-based quantum diagonalization method, have enabled the estimation of ground states for increasingly complex molecules, such as 4Fe–4S involving 54 electrons and 36 orbitals, representing a significant step forward in HPC-QC integration. [62-64]

## Opportunities: Quantum Computing for Energy Materials

While current quantum hardware is not capable of solving materials problems entirely on its own, the integration of QC with classical computing already enabled notable breakthroughs. [3, 15, 29, 65] In these hybrid quantum-classical workflows, each computational paradigm addresses the tasks it handles most effectively: quantum resources are employed for inherently quantum-intensive components such as discrete optimization or Hamiltonian simulation, while classical resources manage tasks like preprocessing, feature engineering, and postprocessing (Figure 2). Such synergistic integration expands the scope of tractable problems, enabling the exploration of vast design spaces that would be computationally challenging using either quantum or classical methods alone.

*Combinatorial Optimization*

Many energy materials discovery are combinatorial optimization problems, involving discrete decisions such as layer sequence, atomic composition, or pixelated patterns. [3, 16] Classical optimization methods (e.g., genetic algorithms, discrete particle swarm optimization) can be trapped in local minima, especially for high-dimensional problems with complex optimization landscapes. [41] The discrete nature of combinatorial optimization also makes gradient-based optimizers not directly applicable. QC, including quantum annealing and QAOA, is naturally suited for such tasks, offering potential quantum speedups. [3, 41]

However, problems need to be mapped into QC-solvable models to take its advantage. Systems that can be represented by Ising model are directly solvable using quantum annealing or QAOA, but such problems are not universal. For general combinatorial optimization problems, one can use data-driven methods to approximate the target problem into a QUBO model. Kitai et al. trained a $2^{nd}$ order factorization machine (FM) model based on data from electromagnetic wave calculations for a thermal radiation metamaterial, and the FM model parameters were used to construct an equivalent QUBO model, whose ground state was identified by quantum annealing. [15] The key advantage of QC leveraged in the study is its capability to efficiently and reliably find the ground state for the given QUBO. A recent benchmark study showed the advantage of QC in finding the ground state of QUBOs derived from practical problems. [41]

However, since the FM model trained on data is a surrogate for the real problem, its accuracy dictates the accuracy of the solution found by QC. To improve the accuracy of the FM model, active learning frameworks are usually employed to gradually approach the true optimal solution of the target problem. This is particularly useful for materials design, where data generation is usually costly and active learning can effectively start from a sparse dataset. A typical active learning scheme leveraging QC is shown in Figure 3A. QC then proposes a promising material candidate based on a given QUBO surrogate. The properties of the candidate are then usually calculated using classical methods, e.g., numerical simulation. The result is then fed back into the FM model in its re-training, and then a new iteration starts. This process iteratively improves the surrogate fidelity and eventually enables convergence toward the global optimal materials.

Notable applications of the QC-enhanced active learning scheme include thermofunctional metamaterials and transparent radiative cooler (TRC, Figure 3B), where their goals are to identify optimal pixelated patterns and multilayer stacks to maximize radiative cooling efficiency. [3, 15] By encoding the material optimization problem into a QUBO through FM surrogate and applying quantum approaches, both works could efficiently explore the vast design space to find optimal structures for radiative cooling. Additionally, Kim et al. experimentally demonstrated the QC-designed metamaterial and proved its energy-saving capabilities (Figure 3C,D). Other examples include thermal emitters and thermophotovoltaics, where QC enabled the discovery of optimal configurations in complex material systems, resulting in desired characteristics. [14, 66] Besides metamaterials, QC-assisted optimization was also demonstrated to guide the optimization of the atomic configurations and compositions of high-entropy alloys to achieve superior mechanical properties (Figure 3E). [16] Beyond materials, QC-assisted optimization can extend to energy systems, including grid layout, storage strategy, and operational scheduling. [67-69]

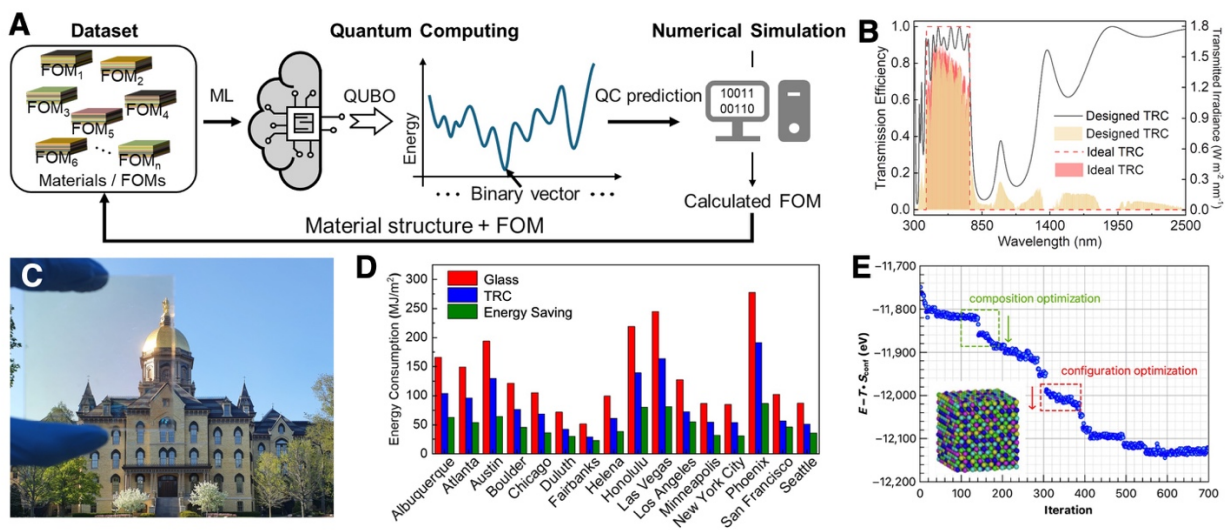

**Figure 3. QC for combinatorial optimization in materials discovery and design. A.** Active learning scheme integrating ML, QC, and numerical simulation in an iterative loop. **B.** Optical characteristics of the QC-designed TRC. **C.** Photograph of the fabricated TRC. **D.** Predicted energy-saving capability of the TRC. Adapted (**A,D**) and reprinted (**B,C**) with permission from [3]. Copyright (2022) American Chemical Society. **E.** Optimization of atomic configurations and composition in high-entropy alloys. Adapted from [16]. CC BY-NC-ND 4.0.

*Hamiltonian Simulation*

QC has shown promise in addressing challenging problems in materials science by directly encoding the quantum nature of molecular systems into quantum circuits. First, a chemically significant property (e.g., the ground-state energy) is formulated into a Hamiltonian. [26, 70] This Hamiltonian is then mapped onto qubits using transformations such as the Jordan-Wigner or Bravyi–Kitaev mappings, which translate the fermionic creation and annihilation operators of the electrons into the Pauli spin operators that QC can process. [26, 34, 71] This "quantum-native" representation enables QC, in principle, to explore exponentially large Hilbert spaces. However, practically exploiting this exponential scaling requires quantum hardware with sufficiently large qubit counts, high gate fidelity, and ultimately fault-tolerant error correction. Once these requirements are satisfied, QC may enable simulations of complex molecular systems that are computationally intractable for classical methods, providing a fundamentally new paradigm for studying strongly correlated and high-dimensional quantum systems. [72]

Ultimately, the process requires moving away from the classical computational paradigm (e.g., DFT), which becomes computationally expensive and inaccurate when applied to strongly correlated or multireference systems, where electrons are highly entangled or multiple electron configurations significantly contribute to the ground state.[13, 26] In classical approaches such as DFT, the complex many-electron wavefunction is replaced by an effective description based on the electron density, which makes the problem tractable but inevitably approximates electron-electron correlations. In contrast, QC can, in principle, directly represent the electronic Hamiltonian, enabling the simulation of complex many-body systems without compromising assumptions if sufficient quantum resources are available.[26, 34] VQE exemplifies this approach by iteratively optimizing a parameterized quantum circuit to minimize the expectation value of the encoded Hamiltonian. Early demonstrations have successfully applied VQE to small molecules, validating the logical mapping from electronic structure problems to qubit-based representations.[38] These developments mark an essential step toward achieving chemically accurate simulations of complex energy materials and accelerating the discovery of new compounds through quantum computation.

A recent study showed quantum simulations of the transition-metal oxide $SrVO_3$ through a combination of compact Wannier representations, hybrid fermion-to-qubit mapping, and efficient circuit compilation.[54] Such advances make it feasible to capture the essential physics of correlated materials on near-term quantum hardware, bringing realistic simulations within reach. Cao et. al. applied quantum embedding combined with chemically intuitive fragmentations to investigate spin polarization in one-dimensional hydrogen chain, the equation of state of two-dimensional hexagonal boron nitride, and magnetic ordering in three-dimensional nickel oxide, illustrating how classically hard problems can be tackled using QC (Figure 4B).[73]

Greene-Diniz et. al. applied QC as a high accuracy post-Hartree Fock solver to simulate $CO_2$ adsorption in metal-organic frameworks (MOFs).[74] By implementing quantum unitary coupled cluster ansatz with single and double excitations in a VQE, they successfully simulated a complex porous system, demonstrating the feasibility of QC-driven material simulations (Figure 4C,D). Furthermore, Li et. al. integrated an adaptive energy sorting approach with the density matrix embedding theory to solve realistic chemical problems.[65] By solving strongly correlated degrees of freedom using quantum algorithms and calculating the remaining parts using classical methods, they analyzed real chemical properties involving strong correlations, such as the reaction energy profile of $C_6H_8$ and the potential energy curve of $C_{18}$.

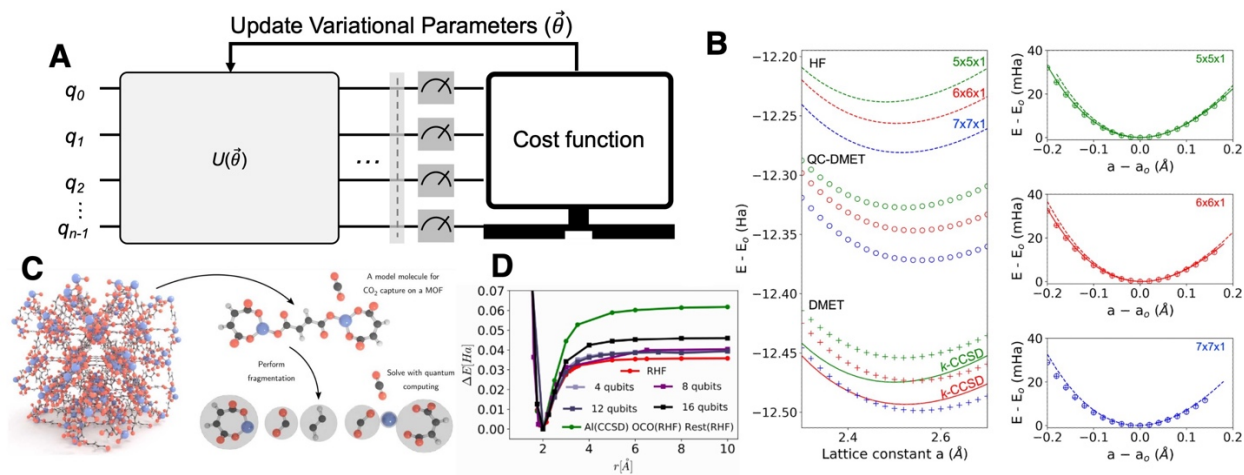

**Figure 4. QC for Hamiltonian simulation in materials simulations. A**. VQAs, a hybrid quantum-classical approach, often used to identify the minimum eigenvalue and its corresponding eigenstate of a Hamiltonian. **B.** Equation-of-state of two-dimensional hexagonal boron nitride calculated using QC. Reproduced from [73]. CC BY 4.0. **C.** The schematic of applying QC to model $CO_2$ capture in metal-organic frameworks. **D.** Dissociation energy $\Delta E$ as a function of Al-$CO_2$ distance ($\gamma$). Reproduced from [74]. CC BY 4.0.

## Challenges and Perspectives

Despite the advances in QC and its increased applications in energy materials, significant challenges remain before its full potential can be realized. These challenges include hardware limitations, ansatz limitations, and difficulties in encoding complex problems into QC-solvable forms. Addressing them will require coordinated efforts across quantum hardware engineering, algorithm development, and domain-specific modeling techniques. Understanding these challenges is crucial for developing strategies that accelerate the adoption of QC in energy materials discovery and related applications.

*Hardware Constraints and Scalability*

**Challenge**: Current quantum hardware poses significant limits on computational scale and fidelity. The number of available qubits is still modest (tens to a few hundreds), coherence times are short (hundreds of microseconds to a few seconds), and gate operations are prone to noise and errors.[27] Limited connectivity between qubits further constrains the complexity of implementable circuits, restricting the depth of quantum operations that can be reliably executed.[23] These limitations are particularly important for high-dimensional materials problems, which often require more than hundreds of qubits and deep circuits to encode realistic material systems. For example, optimizing large-scale material systems (e.g., photonic multilayered structures) requires both a large number of qubits and deep, highly entangled quantum circuits to capture correlation effects accurately. Executing such circuits exceeds the capabilities of most current quantum devices.[29] Additionally, the sampling required to estimate precise expectation values of critical observables introduces overheads that contribute to longer runtimes with the physical limitations of sampling rate varying across quantum hardware.

**Perspective**: Near-term advances in quantum hardware, such as increased number of qubits, incremental reductions in gate error rates, longer coherence times, and improved qubit connectivity, will expand the capabilities of NISQ devices. Techniques such as error mitigation, zero-noise extrapolation, and dynamical decoupling can further extend effective circuit depth and fidelity, enabling meaningful quantum simulations even in the absence of full error correction.[75] In the longer term, the development of fault-tolerant quantum computers with large number of logical qubits will be transformative, allowing scalable simulations of complex material systems at scales.[76] However, we note that realizing fault-tolerant quantum computers will require grouping many physical qubits to encode a single error-corrected logical qubit, leading to a substantial qubit overhead. This physical-to-logical qubit overhead represents a major practical challenge and is expected to play a crucial role in determining both the feasibility and the timeline for achieving large-scale, high-fidelity quantum computing. Distributed strategies leveraging HPC-QC integration provide a practical pathway from the near- to long-term, where problems, circuits, or gate-level operations are decomposed into manageable units on classical HPC systems while quantum resources address the classically intractable sub-instances.[29,77] Collectively, these advances will make large-scale, high-fidelity computations feasible.

*Ansatz Limitations*

**Challenge**: VQAs, including VQE and QAOA, are widely applied across various QC applications. However, they face inherent challenges that limit scalability. One major issue is the phenomenon of barren plateaus, where the optimization landscape becomes extremely flat, making it difficult to identify the optimal variational parameters.[78] Other challenges include limited circuit depth posed by hardware constraints and the difficulty of designing ansatz that are simultaneously expressive, problem-specific, and hardware-compatible.[54] As qubit requirements grow for larger problems, these challenges become more severe, increasing the risk of suboptimal convergence or even preventing convergence entirely, thereby hindering progress toward the ground truth solution.

**Perspective**: Addressing these bottlenecks will require a multifaceted approach. Better ansatz design through physics-informed, problem-specific, or adaptive ansatz can reduce the number of quantum gates required and dimensionality of the optimization space, thereby improving convergence.[38,54,79] ML-assisted circuit design offers a promising direction, where the data-driven models guide optimized quantum circuit constructions.[56] By leveraging prior knowledge, these approaches can help the quantum circuit construction toward the optimized and efficient configurations. Moreover, executing quantum circuits with diverse initial parameters across multiple compute resources in HPC-QC integrated systems allows parallel exploration of the parameter landscape.[79] This parallelization increases the probability of identifying high-quality solutions, accelerating convergence toward the global optimal solution. Together, these strategies may improve the robustness, scalability, and applicability of quantum algorithms to larger and more complex systems.

*Encoding Complex Problems*

**Challenge**: Translating materials optimization problems that involve higher-order interactions into a quantum-compatible representation and solving such complex problems are challenging. Problem encoding typically involves mapping electronic structure, atomic positions, or materials descriptors into qubit states, often requiring ancillary qubits (i.e., additional qubits to help quantum information processing) to map higher-order interaction terms. [24, 80] This process inevitably involves trade-offs between fidelity and resource efficiency. While simplifying these models can reduce qubit requirements or circuit depth, it can also compromise the accuracy. For example, encoding materials optimization problems considering only single- and two-variable interactions can reduce computational overhead but can omit critical many-variable interactions, which can be essential for capturing the full optimization landscape. Similarly, capturing strongly correlated electronic behavior or complex structure–property relationships require large Hamiltonians whose dimensionality can exceed the capability of current QC systems.

**Perspective**: Advances in ML techniques offer a promising route to translate such higher-order problems into quantum-compatible representations. [81] ML models can capture higher-order interactions, which can be directly mapped onto entangled quantum circuits. Quantum circuits can natively encode higher-order correlations through entangled gate operations, enabling more faithful representations of multi-variable interactions. ML is also helpful in identifying the most relevant interaction terms and compressing problem formulations, while quantum hardware directly encodes and processes these complex interactions. QC also enables integrated workflows in which quantum simulation and quantum optimization are coupled in a feedback loop. When the choice of Hamiltonian parameters plays an essential role in the simulated properties, quantum optimization methods can adjust those parameters to steer the Hamiltonian toward a desired simulation outcome. As these approaches mature, QC may enable the solution of complex material systems with high fidelity, advancing the design of high-performance energy materials.

## Summary and Outlook

QC can be transformative for energy materials. By enabling combinatorial optimization of complex systems and simulations of electronic structure, QC provides capabilities that are challenging or intractable for classical methods. Despite these exciting prospects, current quantum hardware limitations, ansatz limitations, and difficulties in problem encoding constrain its practical impact. Noise, limited qubit counts, short coherence times, and connectivity restrictions limit circuit depth and the size of problems that can be realistically addressed. Similarly, quantum algorithms face challenges such as barren plateaus, while translating complex materials problems into qubit-compatible representations often requires trade-offs between accuracy and resource efficiency. Realizing the full potential of QC in energy materials will require the co-development of quantum hardware, quantum algorithms, and domain expertise. Interdisciplinary collaboration among materials scientists, chemists, computer scientists, and quantum engineers is essential to accelerate progress.

Looking forward, the evolution of QC for energy materials can be envisioned across three time horizons (Figure 5):

- Near-term (0-2 years): NISQ devices, coupled with variational algorithms and ML-based surrogate models, enable proof-of-concept demonstrations and small-scale applications in energy materials discovery. Examples include quantum simulations of simple molecules (e.g., $H_2$, $CH_2$, and $H_2O$) and quadratic combinatorial optimization problems (e.g., 1D layered photonic structures, scheduling, and routing). Hybrid quantum-classical methods will capitalize on the complementary strengths of both paradigms: classical computing excels at large-scale data processing, feature engineering, numerical simulations, and resource orchestrations, while quantum computing addresses classically intractable problems such as complex combinatorial search and strongly correlated systems. Implemented on HPC-QC integrated systems, these approaches will expand computational capabilities, supporting more efficient exploration of design spaces.
- Mid-term (2-5 years): Advances in error mitigation, circuit compilation, and HPC-QC integration will extend the reach of quantum computing to larger, more complex material systems. Early-stage implementations of quantum error correction with a modest number of logical qubits will enable more accurate and reliable quantum simulations, allowing the study of larger molecular systems such as $C_6H_6$ and $C_4H_4N_2$ with improved fidelity and complexity. In addition, increased qubit counts will enable higher-order combinatorial optimization and simultaneous optimization of discrete and continuous variables, with potential applications in areas such as 2D metamaterials and diffraction gratings design. Furthermore, tighter integration of ML with HPC-QC platforms will accelerate large-scale energy materials discovery by improving surrogate modeling, parameter initialization, and search efficiency.
- Long-term (> 5 years): Fault-tolerant quantum computing with large numbers of logical qubits will unlock scalable, high-fidelity simulations to enable the design of highly complex systems and ultimately demonstrating quantum advantage. In this regime, QC is expected to accurately treat strongly correlated transition-metal complexes (e.g., Fe(II)-porphyrin complexes, and FeMo-cofactor) whose electronic structures are beyond the reach of classical methods. In addition, large-scale, high-dimensional optimization problems, such as the design of 3D metamaterials and mechanical metamaterials with intricate topologies, will become tractable. Even at this stage, hybrid quantum-classical workflows with HPC-QC integration are expected to remain essential, as QC complements rather than replaces classical approaches.

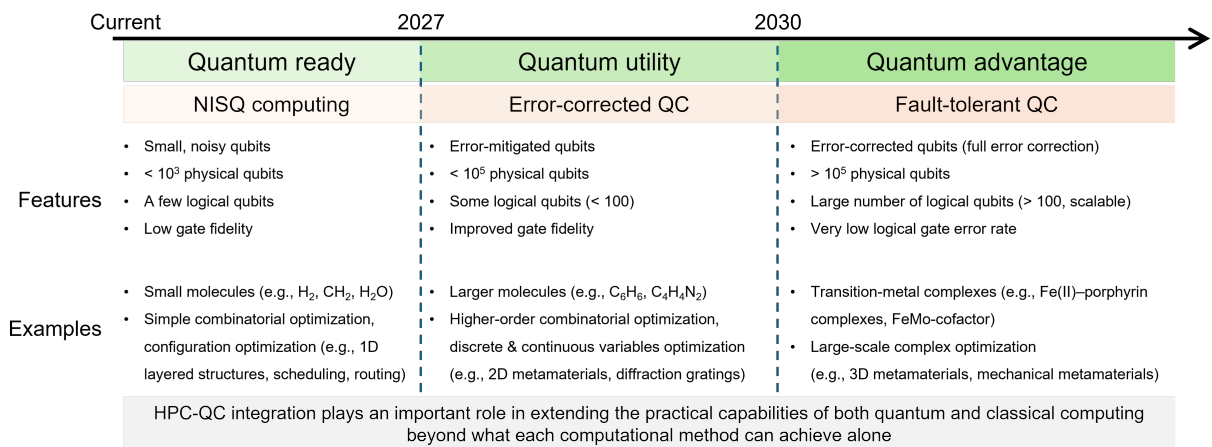

**Figure 5. Roadmap of quantum computing for energy materials,** progressing from near-term (quantum ready), to mid-term (quantum utility), and ultimately long-term (quantum advantage).

These advances promise to accelerate energy materials development, tackle problems that are unsolvable today, and build an interdisciplinary workforce bridging theory, computation, and practical applications. Importantly, quantum computers also have the potential to operate with lower energy consumption and reduced operational costs compared to large-scale classical supercomputers, especially as they scale. [82]

**Highlight 1**. **Common Misconception**: A frequent misunderstanding is that QC will universally replace classical computing. In reality, QC offers exponential or polynomial advantages in a narrow but critical set of problems, including combinatorial optimization, quantum chemistry, and simulation of systems that are inherently "quantum". For most practical problems, however, quantum and classical methods are expected to complement each other.

**Highlight 2. Paradigm Shift in Design**: Integrating QC with ML is already demonstrating practical impacts. By combining the predictive power of ML with the computational advantages of QC, researchers can explore broader design spaces more efficiently, accelerating discovery timelines beyond what classical approaches alone can achieve.

**Highlight 3. Hybrid Quantum-Classical Workflows.** Hybrid quantum-classical workflows will be essential from the near to long term. Even as fault-tolerant quantum computers become available, QC will not universally replace classical computing. Instead, these hybrid frameworks leverage the strengths of each paradigm, maximizing overall computational capability and enabling practical applications in energy materials development.

AUTHOR INFORMATION

**Corresponding Authors**

Seongmin Kim – National Center for Computational Sciences, Oak Ridge National Laboratory; Oak Ridge, Tennessee, 37830, United States. E-mail: kims@ornl.gov

Eungkyu Lee – Department of Electronic Engineering, Kyung Hee University; Yongin-Si, Gyonggi-do, 17104, Republic of Korea. E-mail: eleest@khu.ac.kr

Tengfei Luo – Department of Aerospace and Mechanical Engineering, University of Notre Dame; Notre Dame, Indiana, 46556, United States. E-mail: tluo@nd.edu

**Author**

In-Saeng Suh – National Center for Computational Sciences, Oak Ridge National Laboratory; Oak Ridge, Tennessee, 37830, United States. E-mail: suhi@ornl.gov

Travis S. Humble, Quantum Science Center, Oak Ridge National Laboratory; Oak Ridge, Tennessee, 37830, United States. E-mail: humblets@ornl.gov

Thomas Beck, National Center for Computational Sciences, Oak Ridge National Laboratory; Oak Ridge, Tennessee, 37830, United States. becktl@ornl.gov

**Notes**

The authors declare no competing financial interest.

**Biographies**

**Seongmin Kim** is a Research Scientist in the Quantum-HPC group at Oak Ridge National Laboratory. His research focuses on integrating quantum computing, high-performance computing, and machine learning, with an emphasis on developing scalable, distributed quantum algorithms, to accelerate scientific discovery and optimization of complex systems.

**In-Saeng Suh** is a Senior HPC Quantum System Scientist at the National Center for Computational Sciences, Oak Ridge National Laboratory. His research focuses on integrating quantum computing and high-performance computing for physics and materials science, advancing large-scale simulations through AI-enabled hybrid QC-HPC workflows.

**Travis Humble** is director of the US Department of Energy's Quantum Science Center and a Distinguished Scientist at Oak Ridge National Laboratory. Travis also serves as editor-in-chief for ACM Transactions on Quantum Computing and holds a joint faculty appointment with the University of Tennessee.

**Thomas Beck** is the Section Head for Science Engagement in the National Center for Computational Sciences at Oak Ridge National Laboratory, and is the interim Group Leader of the Quantum-HPC group in NCCS. His research involves computational chemistry and physics, AI, and quantum computing in the energy sciences.

**Eungkyu Lee** is an associate professor in the Department of Electronic Engineering at Kyung Hee University. His research focuses on optimizing light–matter interactions in photonic structures, employing approaches that range from classical optimization theory to state-of-the-art AI-driven techniques and emerging quantum computing algorithms.

**Tengfei Luo** is the Dorini Family Professor and Associate Chair of the Aerospace and Mechanical Engineering Department at University of Notre Dame. He has strong expertise in AI-driven thermal science and materials informatics. He integrates physics-based modeling, machine learning, and quantum optimization for advanced materials, nanoelectronics, and space systems applications.

ACKNOWLEDGMENT

The authors acknowledge that this manuscript has been published in ACS Energy Letters. The published paper is accessible via DOI: https://doi.org/10.1021/acsenergylett.5c04009.

TL and EL would like to acknowledge that this research was supported by the Quantum Computing Based on Quantum Advantage Challenge Research (RS-2023-00255442) funded by the Korea government (MSIT). This research used resources of the Oak Ridge Leadership Computing Facility at the Oak Ridge National Laboratory, which is supported by the Office of Science of the U.S. Department of Energy under Contract No. DE-AC05-00OR22725.

Notice: This manuscript has in part been authored by UT-Battelle, LLC under Contract No. DE-AC05-00OR22725 with the U.S. Department of Energy. The United States Government retains and the publisher, by accepting the article for publication, acknowledges that the U.S. Government retains a non-exclusive, paid up, irrevocable, world-wide license to publish or reproduce the published form of the manuscript, or allow others to do so, for U.S. Government purposes. The Department of Energy will provide public access to these results of federally sponsored research in accordance with the DOE Public Access Plan (http://energy.gov/downloads/doe-publicaccess-plan).